\documentclass[aps,ams,prb,twocolumn,showpacs]{revtex4}
\usepackage{epsfig}
\usepackage{graphicx}  
\usepackage{dcolumn}   
\usepackage{amssymb,amsmath}

\newcommand{\be}{\begin{equation}} \newcommand{\ee}{\end{equation}}
\newcommand{\bea}{\begin{eqnarray}} \newcommand{\eea}{\end{eqnarray}}

\begin{document}

\title{Heat capacity of suspended phonon cavities}

\author{A. Gusso$^{1,2}$\footnote{Present address: Depto. de Ci\^encias Exatas e Tecnol\'ogicas,
Universidade Estadual de Santa Cruz, CEP 45662-000, Ilh\'eus-BA} and Luis G. C. Rego$^1$}
\affiliation{$1-$Departamento de F\'{\i}sica, Universidade Federal de Santa Catarina, \\
88040-900, Florian\'opolis-SC, Brazil
\\ $2-$Departamento de F\'{\i}sica, Universidade Federal do Paran\'a, \\
C.P. 19044, 81531-990 Curitiba-PR, Brazil}

\date{\today}

\begin{abstract}
We present a detailed analysis of the vibrational spectrum and heat capacity of suspended mesoscopic
dielectric plates, for various thickness-to-side ratios at sub-Kelvin temperatures. The vibrational
modes of the suspended cavity are accurately obtained from the three-dimensional (3D) elastic equations
in the small strain limit and their frequencies assigned to the cavity phonon modes. The calculations
demonstrate that the heat capacity of realistic quasi-2D phonon cavities approach the linear dependence
on $T$ at sub-Kelvin temperatures. The behavior is more pronounced for the thinnest cavities, but takes
place also for moderately thick structures, with thickness-to-side ratios $\gamma$=0.1 to 0.2. It is
also demonstrated that the heat capacity of the suspended phonon cavities is invariant under the
product of the temperature (T) with a characteristic lateral dimension (L) of the sample. The present
results establish a lower bound for the heat capacity of suspended mesoscopic structures and indicate
the emergence of the quantum mechanical regime in the dynamics of bounded phonon cavities.
\end{abstract}

\maketitle

\section{Introduction}

Suspended nanostructures have become relevant elements for both basic research and technology.
Micro-electromechanical systems (MEMS), such as cantilevers, gears and membranes, already find
widespread use in several technological applications \cite{handbook}. At the same time, current
developments in surface nanomachining render possible the controlled fabrication of a large variety of
suspended nanostructures \cite{Cleland,Ekinci} having, in particular, an extremely weak thermal
coupling with the environment. As a consequence, ultrasensitive bolometers \cite{Yung02} and
calorimeters \cite{Fon05,Bourgeois05} have been developed with unprecedented sub-attojoule resolution,
for operation in the $T \lesssim 5$ Kelvin temperature range, envisaging the possibility of measuring
the heat capacity of nano-objects and, eventually, even single molecules \cite{Roukes99}. In the realm
of fundamental research, the operation of nanoelectromechanical structures (NEMS) is finally
approaching the quantum regime \cite{Blencowe,Knobel03,Lahaye04}. The construction of suspended solid
state quantum logic gates \cite{Armour02,Cleland04} is amid the applications anticipated for such
structures, since the electron-phonon interaction, which is a source of decoherence and dissipation for
both quantum dot qubits \cite{Gorman05,Hayashi03} and single-electron transistors (SET) \cite{Weig04},
can be controlled in them \cite{Tobias,chaos1,chaos2,Glavin}.

In fact, for most of the cases mentioned above the devices are operated at sub-Kelvin temperatures; the
requirement of ultracold temperatures being specially severe for the operation of quantum logic gates.
For instance, in recent experimental realization,  quantum dot charge-qubits \cite{Gorman05,Hayashi03}
and suspended SET \cite{Weig04} have been operated at a base temperature of 20 mK. It is therefore
reasonable to expect that suspended nanostructures comprising such quantum devices as well as
ultrasensitive bolometers and calorimeters will be functioning at temperatures $T \lesssim 1$K.

Despite the interest, there is not yet a comprehensive theory for the electron-phonon interaction in
suspended nanostructures \cite{Fon02,Qu05}. A central issue of the problem is the difficulty of
rigorously describing the low temperature acoustic phonon spectrum in suspended nano-devices, which is
fundamental for determining {\it (i)} the electron-phonon interaction with its many consequences for
the device operation and {\it (ii)} the device's thermal properties, such as its thermal conductivity
and heat capacity. At sub-Kelvin temperatures, the formalisms adopted for bulk materials may not
produce correct results for the phonon spectrum of suspended nanostructures because the wavelength and
mean free path of the dominant phonons can be bigger than the physical dimensions of the structure.
Moreover, there are no general analytical solutions for the vibrational modes of bounded suspended
plates \cite{Liew95}.

Motivated by these circumstances this work presents a detailed study of the phonon spectrum and the
heat capacity ($C_V$) of suspended rectangular dielectric nanostructures of various thicknesses at
sub-Kelvin temperatures. The vibrational modes of the suspended cavity are accurately calculated from
the three-dimensional (3D) elastic equations in the small strain limit and the obtained frequencies
assigned to the cavity phonon modes. After obtaining a reliable phonon spectrum, with convergence
assured for a few thousand cavity modes, the heat capacity of isolated suspended mesoscopic phonon
cavities having 3D and quasi-2D character is investigated. For such systems, the calculations
demonstrate that the temperature dependence of $C_V$ approaches the linear regime in sub-Kelvin
temperatures, the effect being more pronounced for quasi-2D nanostructures. Nonetheless, a simple model
of plane waves yields a phonon spectrum in good agreement with the 3D elastic model for the very thick
suspended nanostructures. A dimensional analysis of the free vibrational modes also reveal that the
heat capacity of the rectangular phonon cavities has the scale invariant form, that is, $C_V$ is
invariant with respect to the product of the temperature with a characteristic lateral dimension. The
present results indicate that the low temperature heat capacity of quasi-2D suspended nanostructures
may have been underestimated and, therefore, sets a lower bound for their heat capacity.

\section{Theoretical Formulation}

We consider suspended nanostructures of rectangular geometry, with lateral dimensions defined by $L_x$
and $L_y$, and thickness $L_z$. Such a choice is motivated by the fact that several recent experiments
\cite{Yung02,Tighe97, Schwab00}, probing thermal and electrical properties of suspended nanostructures,
have utilized square or rectangular plates with thickness-to-side ratios ($\gamma = L_z/L_y$) including
0.1 [\onlinecite{Tighe97}], 0.04 [\onlinecite{Schwab00}], and 0.015 [\onlinecite{Yung02}]. The first
structure is considered to be a moderately thick plate, while the last two cases  are examples of thin
plates. Because the suspended nanostructures are usually made of non-crystalline materials, like
poly-silicon and amorphous SiN (Silicon Nitride), in this work the phonon cavities are taken to be
homogeneous and isotropic rectangular structures.

Because at sub-Kelvin temperatures the dominant phonon modes are long wavelength acoustic ones, with a
mean free path that exceeds the dimensions of the structure \cite{Yung02,Tighe97, Schwab00}, we resort
to the elasticity theory to obtain the phonon spectrum of the  suspended nanostructures. In this limit
the phonons correspond to the free vibrational modes of the cavity \cite{chaos2}, as determined by the
elastic theory of solids \cite{Graff}. The continuum elasticity model has been successfully used to
describe the properties of propagating phonons in beams \cite{beams,Santamore}, thin membranes
\cite{Tobias,slabs}, and arrays of nanomechanical resonators \cite{Photiadis,Zalalutdinov}.

To secure the correct description of the thermal properties of suspended nanostructures of a few $\mu
m^2$ in area in the sub-Kelvin temperature regime, at least a few thousand vibrational modes have to be
calculated with confidence. A variety of methods intended to calculate the free vibrations of thick
plates have been developed \cite{Liew95}. Since the simplified models like the Classical Plate Theory
(CPT) \cite{Leissa} are adequate only for the lowest modes of thin plates, a three-dimensional analysis
of the free vibrations of the cavity is necessary. In general such methods utilize the Rayleigh-Ritz
formalism to determine the displacement field, which is represented as a series of orthogonal
polynomials \cite{Liew,Zhou}. In this work we follow the procedure developed by Zhou et al. \cite{Zhou}
due to its simplicity and generality as well as for producing very accurate natural frequencies. For
the sake of completness, the method is summarized next.

For the problem of the free vibrations of isotropic structures in the small strain approximation, the
kinetic ($\textsf{T}$) and strain ($\textsf{U}$) elastic energy functionals for the displacement field
$\mathbf{u}(\vec{r},t)=\mathbf{U}(\vec{r})e^{i\omega t}$ can be written as
\begin{eqnarray}
\textsf{T} &=& \frac{\rho}{2} \int \left[ \left(\frac{\partial u_x}{\partial t}\right)^2 +
\left(\frac{\partial u_y}{\partial t}\right)^2 + \left(\frac{\partial u_z}{\partial t}\right)^2 \right] dv \label{T} \\
\textsf{U} &=& \frac{E}{2(1+\nu)} \int \left(\frac{\nu \Lambda^2_1}{1-2\nu} + \Lambda_2 +
\frac{\Lambda_3}{2} \right) dv \ , \label{V}
\end{eqnarray}
where $\rho$ is the mass density, $E$ is the Young's modulus and $\nu$ the Poisson's ratio. The
$\Lambda$ quantities in the strain energy term are: $\Lambda_1= \sum_i \varepsilon_{ii}$, $\Lambda_2 =
\sum_i \varepsilon_{ii}^2$ and $\Lambda_3 = \sum_{i<j} \varepsilon_{ij}$, for $i,j=(x,y,z)$, with the
components of the strain $\varepsilon_{ii} =\partial_i u_i$ and $\varepsilon_{ij} =
\partial_j u_i + \partial_i u_j$. It is convenient to normalize the coordinates
with respect to the dimensions of the plate, defining the dimensionless variables $\xi=2x/L_x$,
$\eta=2y/L_y$ and  $\zeta=2z/L_z$ in the interval [-1,1]. The time-independent displacement field
$\mathbf{U}(\vec{r})$ is then written as a sum of orthogonal Chebyshev polynomials multiplied by
boundary functions $F_{\delta}(\xi,\eta)$
\begin{eqnarray}
U_{\mathrm{x}}(\xi,\eta,\zeta) &=& F_\mathrm{x}(\xi,\eta) \sum_{i,j,k} A_{ijk}P_i(\xi)P_j(\eta)P_k(\zeta) \\
U_{\mathrm{y}}(\xi,\eta,\zeta) &=& F_\mathrm{y}(\xi,\eta) \sum_{i,j,k} B_{ijk}P_i(\xi)P_j(\eta)P_k(\zeta) \\
U_{\mathrm{z}}(\xi,\eta,\zeta) &=& F_\mathrm{z}(\xi,\eta) \sum_{i,j,k}
C_{ijk}P_i(\xi)P_j(\eta)P_k(\zeta) \ , \label{displacements}
\end{eqnarray}
with the summations beginning from zero. The functions $P_n(\chi)$ are Chebyshev polynomials of the
first kind and degree $n$, defined by the relation
\begin{eqnarray}
P_n(\chi) = \cos\left[n\arccos(\chi)\right] \ ,\label{Chebyshev}
\end{eqnarray}
with $n$ a non-negative integer. The boundary functions $F_\mathrm{x}(\xi,\eta)$,
$F_\mathrm{y}(\xi,\eta)$ and $F_\mathrm{z}(\xi,\eta)$ have the general form $F_{\delta}(\xi,\eta) =
f^1_{\delta}(\xi)f^2_{\delta}(\eta)$, with $\delta=\mathrm{x},\mathrm{y},\mathrm{z}$. For our purposes
the boundary conditions of interest are FF (free-free), CC (clamped-clamped) and CF/FC, which
correspond, respectively, to the functions $f^1_{\delta}(\xi) \equiv f^2_{\delta}(\eta) = 1$,
$f^1_{\delta}(\xi) \equiv f^2_{\delta}(\eta) = 1-\eta^2$, and $f^1_{\delta}(\xi) \equiv
f^2_{\delta}(\eta) = 1 \pm \eta$.

Substituting the series representation for $\mathbf{u}(\vec{r},t)=\mathbf{U}(\vec{r})e^{i\omega t}$
into Eqs. (\ref{T}) and (\ref{V}), one obtains $\textsf{T}_{max}$ and $\textsf{U}_{max}$, which are the
maximum values of $\textsf{T}$ and $\textsf{U}$ during a vibratory cycle. The frequency determinant is
formulated by minimizing the functional $\textsf{U}_{max}-\textsf{T}_{max}$ with respect to each of the
coefficients $\{A\}$, $\{B\}$ and $\{C\}$, to produce the 3D elastic equations of motion
\begin{eqnarray}
\left(\left[K\right] - \Omega^2\left[M\right]\right) \left(%
\begin{array}{c}
  \left\{A\right\} \\
  \left\{B\right\} \\
  \left\{C\right\} \\
\end{array}%
\right) = 0 \ , \label{eigen}
\end{eqnarray}
where $\Omega = \omega L_x \sqrt{\rho/E}$ is a dimensionless parameter and $\omega$ is the free
vibration frequency, to be assigned to the phonons. In Eq. (\ref{eigen}), $\left[K\right]$ and
$\left[M\right]$ denote the symmetric stiffness matrix and the block diagonal mass matrix,
respectively, which can be found in explicit form in Ref. [\onlinecite{Zhou}]. Another useful
dimensionless parameter associated with the frequency is $\Delta =
\left(\Omega/\lambda\gamma\pi^2\right)\sqrt{12(1-\nu^2)}$, with $\lambda = L_x/L_y$ and $\gamma =
L_z/L_y$, which yields the normalized frequencies for the family of all rectangular plates with the
same aspect ratios ($\lambda$ and $\gamma$) and elastic constants ($E$ and $\nu$).

An important aspect of the present analysis is the reliability of the phonon spectrum to be used in the
evaluation of the thermodynamical properties of the nanostructure. For that reason the convergence of
the highest frequencies was limited to be within 5\%. The spectrum span can be extended by increasing
the amount of basis functions $P_n(\chi)$ used in the representation of $\mathbf{U}(\vec{r})$. In the
instance of the thickest square plate to be considered, with $\gamma = 0.5$, we have used $n_x=n_y=29$
and $n_z=13$ Chebyshev polynomials, yielding approximately 12000 reliable modes. For the thinnest plate,
$\gamma = 0.02$, the best results were obtained for $n_x=n_y=51$ and $n_z=4$ that yielded 4000 reliable
phonon modes. The number of reliable frequency modes will determine the maximum temperature ($T_{\rm
max}$) for which the heat capacity can be calculated with confidence.

\begin{table}
\caption{\label{constants} Physical constants of relevant materials in the amorphous phase.}
\begin{tabular}{c c c c}
\hline \hline
Material \ \ \ \ \ \ & $\rho$(g/cm$^3$)  \ \ \ \ \ \ &  $E$(GPa) \ \ \ \ \ \ \ \ \ \ & $\nu$ \\
GaAs \ \cite{aGaAs} & 5.1 & 71 \ \ \ \ \ \ & 0.32 \\
Si \ \cite{aSi} & 2.3 & 170 \ \ \ \ \ \ \ & 0.22 \\
SiN \ \cite{SiN} & 3.1 & 285 \ \ \ \ \ \ & 0.20  \\
SiC \ \cite{SiC} & 3.0 & 400 \ \ \ \ \ \ & 0.20  \\
\hline \hline
\end{tabular}
\end{table}

The dependence of the phonon frequencies on the material parameters is such that higher frequencies are
obtained for stiff and light materials. Table \ref{constants} contains the values of the mass density,
Young's modulus and Poisson's ratio for materials of relevance for the fabrication of NEMS. This work
investigates nanostructures made of amorphous silicon carbide (a-SiC) because of its high rigidity and
widespread use in the fabrication of suspended NEMS.

\begin{figure}[h]
\includegraphics[width=8cm]{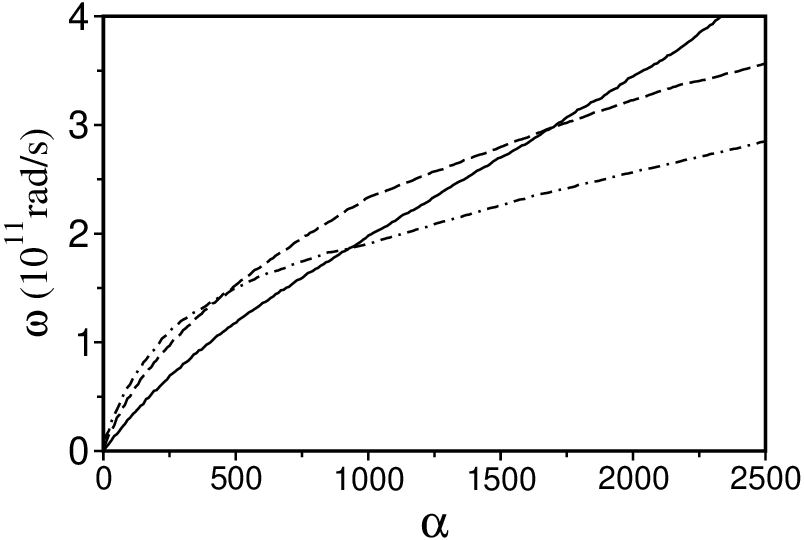}
\caption{Natural vibration frequencies, as a function of the mode index $\alpha$, calculated by the 3D
method for a free standing square ($\lambda = 1$) a-SiC nanostructure of sides $L = 2 \mu m$ and
$\gamma=L_z/L$ equal to 0.02 (solid), 0.05 (dashed) and 0.1 (dot-dashed).} \label{spectra}
\end{figure}

The calculated spectra exhibit significant dependence on the dimensions of the nanostructure, as
illustrated in Figure \ref{spectra} for the first 2500 vibrational modes of free standing a-SiC
mesoscopic structures. The structures have the same lateral dimensions $L_x=L_y=L=2\ \mu m$ but
different thickness-to-side ratios: $\gamma$= 0.02 (solid), 0.05 (dashed) and 0.1 (dot-dashed). At the
lower part of the spectrum the frequencies are higher for the thick plates, however, the behavior is
reversed as the mode index $\alpha$ increases. The frequencies are also inversely proportional to the
area of the plate. Moreover, by a numerical analysis of the eigenfrequencies it was observed that the
vibrational spectrum of the cavities can be very well described by the form $\omega = \omega_0
\alpha^{\phi}$ in two limiting cases: for the quasi-2D phonon cavities ($\gamma \leq 0.02$) the fitting
yields $\phi \lesssim 1$, whereas  for the three-dimensional (thick) phonon cavities ($\gamma
> 0.2$) one obtains $0.4 < \phi < 0.5$ . Between the two cases, {\it i.e.} for moderately thick
nanostructures, the frequencies cannot be well described by a single power curve.
\\
\\

In the following we examine the heat capacity of the phonon cavities, as predicted by the 3D analysis.
For the sake of comparison, we also utilize a basic model to describe the confined phonons, which
comprises some of the reductionist features commonly found in the literature \cite{Roukes99,Anghel99}.
Its main assumption is that the phonons can be described by plane waves with three independent
polarizations: one longitudinal and two transverse. In addition, as the dimensions of the structure
become sufficiently small, {\it i.e.} comparable to the mean free path of the phonons, these become
standing waves satisfying the appropriate boundary conditions.  The method is here designated Bounded
Plane Wave Model (BPWM). For the nanostructures under consideration and because of the very low
temperature, it is assumed that the phonons form standing waves in all three directions. In comparison
with the 3D analysis, different predictions for the specific heat are expected on the basis of the
BPWM, owing to its naive representation of the phonon modes in suspended nanostructures. However,
despite the simplicity, it will be shown that the BPWM can describe the heat capacity of thick phonon
cavities quite well.

According to the BPWM, the phonon spectrum of a freely suspended nanostructure is easily obtained from
the wave vectors
\begin{eqnarray} \kappa_{lmn}^2 = \pi^2 \left[ \left(
\frac{l}{L_x} \right)^2 +\left( \frac{m}{L_y} \right)^2 +\left( \frac{n}{L_z} \right)^2 \right]\ ,
\end{eqnarray}
with $l$, $m$ and $n$ integers. The frequencies for the longitudinal and transverse modes are given by
$\omega^l_{lmn} = v^l \kappa_{lmn}$ and $\omega^t_{lmn} = v^t \kappa_{lmn}$, with the sound velocities
of the a-SiC obtained from the elastic constants of the material: $v^l = 12,170$ m/s and $v^t = 7,450$
m/s.

\section{Heat Capacity of Suspended Phonon Cavities}

Having calculated the displacement modes $\mathbf{U}_\alpha(\vec{r})$ and the associated
eigenfrequencies $\omega_\alpha$, corresponding to the free vibrations of the plate, the quantum
mechanical phonon modes of the cavity are obtained by the standard quantization procedure
\cite{chaos2}. As a result, we ascribe the energy  $\mathcal{U} = \sum_{\alpha} (n_\alpha +1/2) \hbar
\omega_\alpha$ to the phonon system and calculate the quantum mechanical heat capacity of the phonon
cavity as
\begin{eqnarray} C_V(T)
= \frac{\partial \cal{U}}{\partial T} = \frac{\partial}{\partial T} \sum_\alpha
\frac{\hbar\omega_\alpha}{\exp{(\hbar \omega_\alpha/k_BT)}-1} \ , \label{CT}
\end{eqnarray}
with $n_\alpha$ given by Planck's distribution. Since the harmonic regime complies with the small
strain limit that is assumed in the present derivations, the constant-volume specific heat ($c_v$) must
equal the constant-pressure specific heat ($c_p$) \cite{Ashcroft}. Moreover, for bulk metallic samples
it is generally found that the low temperature specific heat varies as $c_v = AT + BT^3$, comprising
the electron and phonon contributions, respectively.

\begin{figure}[h]
\includegraphics[width=6.5cm]{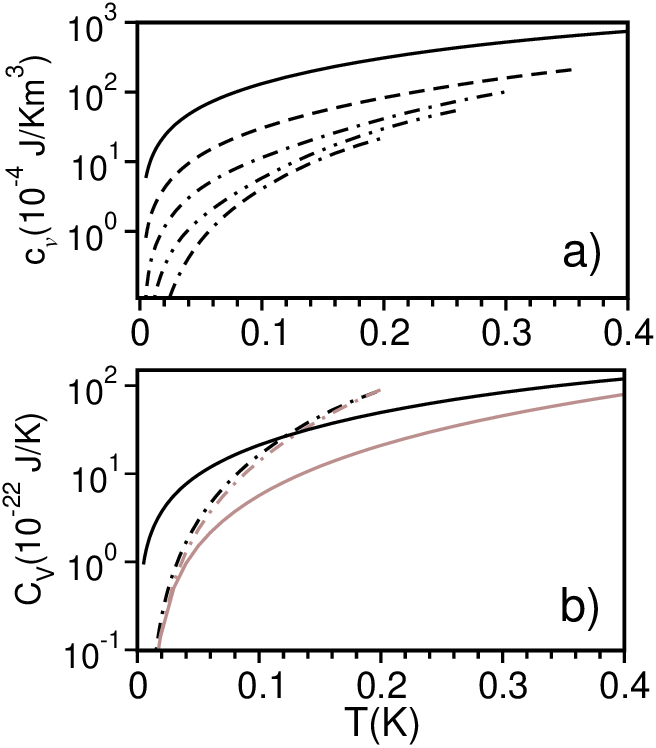}
\caption{a) Specific heat ($c_v = C_V/V$) as a function of temperature obtained from the 3D analysis
for a free standing a-SiC square cavity of lateral dimensions $L= 2 \ \mu m$ and $\gamma$ = 0.02, 0.05,
0.1, 0.2 and 0.5, in that order from top to bottom. b) The heat capacity ($C_V$) for the quasi-2D
$\gamma$ = 0.02 (solid) and fully 3D $\gamma$ = 0.5 (dashed). The gray curves are predictions from the
BPWM (refer to the text).} \label{C_2mu}
\end{figure}

Next we show predictions for the heat capacity obtained through the 3D analysis as well as results
gained by the simplified method. The phonon cavities to be considered are free standing a-SiC square
plates ($\lambda = 1$) with lateral dimensions $L = 2\ \mu m$ and having different thickness-to-side
ratios, namely $\gamma$ = 0.02, 0.05, 0.1, 0.2 and 0.5. $C_V$ is calculated for temperatures $T
\leqslant T_{\rm max}$, where $T_{\rm max}$ is the maximum temperature that allows reliable results to
be obtained with the available phonon modes. That is, if $T > T_{\rm max}$ additional modes must be
included in the calculation of $C_V$ since the occupation of the high energy modes is increases. Figure
\ref{C_2mu}(a) presents the specific heat $c_v = C_V/V$ as a function of temperature for phonon
cavities of different thicknesses, as obtained through the 3D analysis. For temperatures $T \lesssim
100$ mK the calculations reveal more than an order of magnitude difference between the quasi-2D
($\gamma \leqslant 0.05$) and the three-dimensional ($\gamma \gtrsim 0.2$) phonon cavities. Figure
\ref{C_2mu}(b) shows the heat capacity ($C_V(T)$) for two limiting cases, represented by the quasi-2D
($\gamma = 0.02$) and fully three-dimensional ($\gamma = 0.5$) suspended nanostructures. Results gained
from the BPWM are also shown by the gray curves. Particularly, the BPWM predicts very good results for
the three-dimensional cavities, but seriously underestimates the heat capacity of the quasi-2D
structures.

\begin{figure}[h]
\includegraphics[width=8cm]{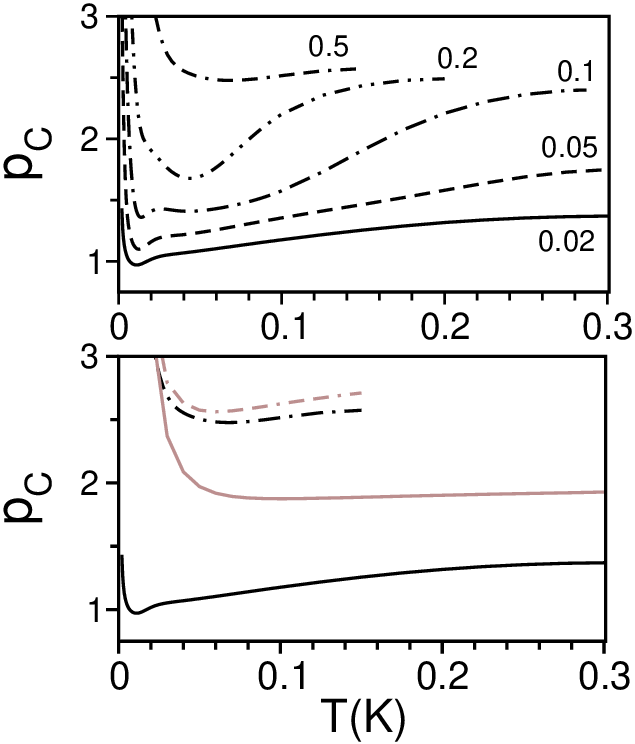} \caption{$p_C(T)$ for free standing a-SiC square plates
with $L= 2\ \mu$m. In the upper panel the predictions of the 3D analysis for different $\gamma$, as
indicated by the labels. In the lower panel $p_C$ for the quasi-2D ($\gamma$ = 0.02) and fully 3D
($\gamma$ = 0.5) cases. The gray curves are the results gained from the BPWM.} \label{pC_fig}
\end{figure}

Throughout the analysis we have considered thin as well as thick suspended nanostructures. That raises
the question, how is the system's dimensionality reflected on the behavior of $C_V(T)$? In Ref.
[\onlinecite{Anghel99}] the dependence of the specific heat on the dimensionality of the system was
investigated with a model similar to the BPWM. It was shown that the relation $C_V \propto T^d$ should
hold for a confined phonon gas in the low temperature limit, with $d$ as the system's dimensionality,
supporting the inaccurate notion that $C_V \propto T^2$ for quasi-2D phonon cavities at sub-Kelvin
temperatures. Here we perform such an analysis and demonstrate instead that the heat capacity of
realistic quasi-2D phonon cavities approaches the linear dependence $C_V \propto T$ in the low
temperature limit. For that purpose consider the quantity
\begin{eqnarray}
p_C(T) = T \frac{\partial (\ln C_V(T))}{\partial T} \ ,
\end{eqnarray}
which provides the temperature dependence of $C_V(T)$. For instance, if the heat capacity is given by
$C_V \propto T^{\alpha}$, we have simply $p_C = \alpha$.

Figure \ref{pC_fig} presents the calculated values of $p_C(T)$ for free standing square plates with
lateral dimensions $L = 2 \mu m$ and the thickness-to-side ratios previously considered, ranging
through the quasi-2D to the fully 3D cases. The upper panel shows that $C_V$ approaches the linear $C_V
\propto T$ behavior at temperatures $T \lesssim 0.3\ K$, particularly in the case of the thinnest
cavities with $\gamma$ = 0.02 and 0.05. Such a result is expected to hold for strict 2D systems like
graphene, although the effect has been predicted also for the specific heat of a 2D array of
nanomechanical resonators \cite{Photiadis}. In fact, the sub-$T^2$ behavior is observed here even in
the case of the moderately thick cavities with $\gamma$ = 0.1 and 0.2, at lower temperatures. As the
temperature increases $p_C$ tends to 3, indicating that the thickness of the cavity becomes much larger
than the dominant phonon wavelength; on the other hand, for vanishing small temperatures, lower than
the fundamental vibrational energies, the phonon cavity behaves as a 0D system. In this case the
specific heat decreases exponentially and $p_C$ diverges with $T^{-1}$. The $C_V \propto T$ behavior is
commonly associated with quasi-1D systems like single-wall nanotubes (SWNT) at low temperatures
\cite{SWNT}. However, as the temperature decreases beyond $T \lesssim 0.5 K$ a sublinear behavior
is observed n such systems \cite{Lasjaunias}. The effect is ascribed to the overwhelming contribution of the flexural
modes to $C_V$, since those modes present $\omega(q) \propto q^2$ and consequently $C_V \propto
T^{1/2}$ [\onlinecite{Glavin01,Popov}].

The lower panel of Figure \ref{pC_fig} compares the results obtained from the 3D analysis with those of
the simple BPWM. According to earlier calculations, both methods produce similar results for fully
three-dimensional structures, but the BPWM yields the wrong $p_C \thickapprox 2$ value for quasi-2D
structures. In the last case, it has been verified that the simple classical plate theory (CPT) for
flexural modes, which reduces the dimensions of the problem from three to two by incorporating some of
the plate's characteristics such as bending moments \cite{Leissa,Graff}, yields a close estimate for
the temperature dependence of $C_V(T)$. The CPT fails, however, as the temperature raises above
$T\gtrsim \hbar \pi v_s/(L_z k_B)$, $v_s$ being the sound velocity, because longitudinal and torsional
modes begin to contribute significantly to $C_V$. Because of the significant difference between the
predictions made by the 3D analysis and the BPWM for thin nanostructures, $p_C$ may be a convenient
observable to experimentally determine the emergence of coherent quantum mechanical dynamics in mesoscopic phonon
cavities.

An additional property of the suspended phonon cavities is the scale invariant character of their heat
capacity, described as $C_V = \mathcal{F}(TL)$, where $\mathcal{F}$ represents the functional in Eq.
(\ref{CT}). Namely, $C_V$ is invariant regarding the product of the temperature ($T$) with a
characteristic lateral dimension ($L$) of the structure.
 For the sake of clarity we consider a square cavity, but
the same result can be derived for rectangular cavities with $L_x = \lambda L_y$, or triangular ones.
First notice that the phonon frequency can be written as
\begin{eqnarray}
\omega_\alpha = \frac{\pi^2 \gamma}{L} \ \Delta_\alpha \sqrt{\frac{E}{12\rho(1-\nu^2 )}} \
.\label{omega}
\end{eqnarray}
Thus $\omega_\alpha \propto 1/L$, for plates of a given thickness-to-side ratio $\gamma$. The
dimensionless parameter $\Delta_\alpha$ is also a function of $\gamma$ and $\lambda$, therefore
independent of the absolute dimensions of the plate. Then, from the definition of the heat capacity,
Eq. (\ref{CT}), with the derivative and summation operations commuted, it is easily verified that a
transformation that leaves $\omega_\alpha/T \propto 1/(LT)$ invariant does not change the heat
capacity. Consequently, the results that have been presented for square phonon cavities of lateral
dimension $L = 2 \mu m$ can be generalized for congruent cavities of arbitrary size $L'$, with the
temperature re-scaled to $T' = (L/L')T$.

Different types of suspended cavities were also investigated, such as bridges (CCFF) and
cantilever-like (CFFF) structures, yielding results in qualitative agreement with those previously
illustrated, for both the vibrational spectrum and the specific heat. It is observed that the parameter
$p_C$ shows a tendency towards the value 1 for elongated structures. For instance, in the case of
$\gamma = 0.1$ and $\lambda = L_x/L_y = 4$ we obtained $p_c \thickapprox 1.5$ at $T = T_{\rm max}$.

The hitherto calculations of the heat capacity of phonon cavities have not taken into account the
additional degrees of freedom  comprised by impurities, disorder and surface defects, etc. that will be
responsible for an increase of $C_V$. In particular, the specific heat of bulk noncrystalline solids
exhibits an anomalous linear variation with the temperature  \cite{Zeller} for $T <$ 1 K. The present
results, however, set a lower bound for the specific heat of such dielectric nanostructures.

\section{Conclusions}

We presented a detailed investigation of the vibrational spectrum and the heat capacity of suspended
dielectric mesoscopic structures of various thicknesses at sub-Kelvin temperatures. More than 4000
frequency modes of the cavity were accurately obtained from the 3D elastic equations in the small
strain regime. It is therefore demonstrated that the low temperature heat capacity of realistic
quasi-2D phonon cavities have an approximate linear dependence on $T$, a result that contradicts
estimates obtained by simple models. The sub-$T^2$ variation of the heat capacity is observed even for
the moderately thick mesoscopic structures. The results show the importance of a fully 3D analysis
based on the elastic equations of suspended plates, bridges and cantilevers, for the correct
determination of their thermal properties. Finally, the sub-$T^2$ effect evidences the quantum
mechanical nature of the phonon cavity dynamics and sets a lower bound for their specific heat. The
reported results should have special interest for suspended nanostructures intended to be part of solid
state quantum devices.

\section{Acknowledgments}

The authors acknowledge financial support from CNPq/Brasil and funding provided by {\it Projeto
Universal} - CNPq. We thank W. Figueiredo and M.E.G. da Luz for comments and suggestions.



\begin{thebibliography}{10}

\bibitem{handbook} B. Bushan (editor), {\it Springer Handbook of Nanotechnology} (Springer,
Berlin,2004).

\bibitem{Cleland} A. N. Cleland, {\it Foundations of Nanomechanics} (Springer-Verlag, 2002).

\bibitem{Ekinci} K.L. Ekinci and M.L. Roukes, Rev. Sc. Inst. {\bf 76}, 061101 (2005).

\bibitem{Yung02} C. S. Yung, D. R. Schmidt, and A. N. Cleland, Appl. Phys. Lett.
{\bf 81}, 31 (2002).

\bibitem{Fon05} W. Chung Fon, Keith. C. Schwab, John M. Worlock, and Michael L.
Roukes, Nano Lett. {\bf 5}, 1968 (2005).

\bibitem{Bourgeois05} O. Bourgeois, S.E. Skipetrov, F. Ong, and J. Chaussy, Phys. Rev. Lett. {\bf 94},
057007 (2005).

\bibitem{Roukes99} M. L. Roukes, Physica B {\bf 263-264}, 1 (1999).

\bibitem{Knobel03} R.G. Knobel and A.N. Cleland, Nature {\bf 424}, 291 (2003).

\bibitem{Lahaye04} M.D. LaHaye, O. Buu, B. Camarota, K.C. Schwab, Science {\bf 304}, 74 (2004).

\bibitem{Blencowe} M. Blencowe, Phys. Rep. {\bf 395}, 159 (2004).

\bibitem{Armour02}
A. D. Armour, M. P. Blencowe, and K. C. Schwab, Phys. Rev. Lett. {\bf 88}, 148301 (2002).

\bibitem{Cleland04}
A.N. Cleland and M.R. Geller, Phys. Rev. Lett. {\bf 93}, 070501 (2004).


\bibitem{Hayashi03} T. Hayashi, T. Fujisawa, H.D. Cheong, Y.H. Jeong, and Y. Hirayama, Phys. Rev.
Lett. {\bf 91}, 226804 (2003); T. Fujisawa, T. Hayashi and Y. Hirayama, J. Vac. Sci. Tech. B {\bf 22},
2035 (2004).

\bibitem{Gorman05} J. Gorman, E.G. Emiroglu, D. G. Hasko, and D. A. Williams, Phys. Rev. Lett. {\bf 95},
090502 (2005).

\bibitem{Weig04} E.M. Weig {\it et al.}, Phys. Rev. Lett. {\bf 92}, 046804 (2004).

\bibitem{Tobias} S. Debald, T. Brandes, and B. Kramer, Phys. Rev. B {\bf 66}, 041301(R) (2002).

\bibitem{chaos1} L. G. C. Rego, A. Gusso, and M. G. E. da Luz, J. Phys. A: Math. Gen.
{\bf 38}, L639 (2005).

\bibitem{chaos2} A. Gusso, M. G. E. da Luz, and L. G. C. Rego, Phys. Rev. B {\bf 73}, 035436 (2006).

\bibitem{Glavin} B.A. Glavin, V.I. Pipa, V.V. Mitin, and M.A. Stroscio, Phys. Rev.B {\bf 65}, 205315 (2002).

\bibitem{Fon02} W. Fon, K.C. Schwab, J.M. Worlock, and M.L.Roukes, Phys. Rev. B {\bf 66}, 045302, (2002); S.
Barman and G.P. Srivastava, Phys. Rev. B {\bf 73}, 205308 (2006).

\bibitem{Qu05} S.X. Qu, A.N. Cleland, and M.R. Geller, Phys. Rev. B {\bf 72}, 224301 (2005).

\bibitem{Liew95} A comprehensive literature review of the problem can be found in: L.M. Liew, Y. Xiang,
and S. Kitipornchai, J. Sound and Vib. {\bf 180(1)}, 163 (1995).

\bibitem{Tighe97} T. S. Tighe, J. M. Worlock, and M. L. Roukes, Appl. Phys.
Lett. {\bf 70}, 2687 (1997).

\bibitem{Schwab00} K. Schwab et al., Nature {\bf 404}, 974 (2000).

\bibitem{Graff} Karl F. Graff, {\it Wave Motion in Elastic Solids}, (Dover, NY, 1975).

\bibitem{beams} L. G. C. Rego and G. Kirczenow, Phys. Rev. Lett. {\bf 81}, 232 (1998).

\bibitem{Santamore} D.H. Santamore and M.C. Cross, Phys. Rev. B {\bf 66}, 144302 (2002).

\bibitem{slabs} T. K\"uhn, D.V. Anghel, J.P. Pekola, M. Manninen, and Y.M. Galperin, Phys. Rev. B {\bf 70}, 125425 (2004).

\bibitem{Photiadis} D.M. Photiadis, J.A. Bucaro, and X. Liu, Phys. Rev. B {\bf 73}, 165314 (2006).

\bibitem{Zalalutdinov} M.K. Zalalutdinov et al., Appl. Phys. Lett. {\bf 88}, 143504 (2006).

\bibitem{Leissa} G. F. Elsbernd and A. W. Leissa,
{\it Developments in Theoretical and Applied Mechanics} {\bf 4}, 19 (1970).

\bibitem{Zhou} D. Zhou, Y.K. Cheung, F.T.K. Au, and S.H. Lo, Int. J. of Solids Structures {\bf
39}, 6339 (2002).

\bibitem{Liew} K.M. Liew, K.C. Hung, and M.K. Lim, Int. J. Solids Structures {\bf 30}, 3357 (1993);
J. Appl. Mech. {\bf 62}, 159 (1995).

\bibitem{aGaAs} M. C. Rigdway, C. J. Glover, G. J. Foran, and K. M. Yu, J. Appl. Phys.
{\bf 83}, 4610 (1998); \\
Ingvar Ebbsj\"o {\it et al.}, J. Appl. Phys. {\bf 87}, 7708 (2000).

\bibitem{aSi} W. N. Sharpe Jr., B. Yuan, R. Vaidyanathan, and R. L. Edwards, in
Proceedings of the 10th IEEE  International Workshop on Microelectromechanical Systems,
 424 (1997).

\bibitem{SiN}  A. Khan, J. Philip, and P. Hess, J. Appl. Phys. {\bf 95}, 1667 (2004).

\bibitem{SiC} M. A. El Khakani, M. Chaker, M. E. O´Hern, and W. C. Oliver, J. Appl. Phys. {\bf 82}, 4310
(1997); \\
R. F. Wiser, M. Tabib-Azar, M. Mehregany, and C. A. Zorman, J. Microelectromech. Syst. {\bf 14}, 579
(2005).

\bibitem{Anghel99} D. V. Anghel and M. Manninen, Phys. Rev. B {\bf 59}, 9854 (1999).

\bibitem{Ashcroft} N.W. Ashcroft and N.D. Mermin, {\it Solid State Physics} (Saunders College,
Philadelphia, 1976).


\bibitem{SWNT} W. Yi, L. Lu, Z. Dian-lin, Z.W. Pan, and S.S. Xie, Phys. Rev. B {\bf 59}, R9015 (1999); J. Hone, B. Batlogg, Z. Benes, A.T. Johnson, and J.E. Fischer, Science {\bf 289}, 1730
(2000).

\bibitem{Lasjaunias} J.C. Lasjaunias, K. Biljakovic, Z. Benes, J.E. Fischer, and P. Monceau, Phys. Rev.
B {\bf 65}, 113409 (2002).

\bibitem{Glavin01} B.A. Glavin, Phys. Rev. Lett. {\bf 86}, 4318 (2001).

\bibitem{Popov} V.N. Popov, Phys. Rev. B {\bf 66}, 153408 (2002).


\bibitem{Zeller} R.C. Zeller and O. Pohl, Phys. Rev. B {\bf 4}, 2029 (1971).



\end{thebibliography}
\end{document}